# ENTROSTAT

Viktor I. Shapovalov

The Volgograd Branch of Moscow Humanitarian-Economics Institute, Volgograd, Russia.

shavi@rol.ru

The reasons for introducing the concept of the entrostat in statistical physics are examined. The introduction of the concept of the entrostat has allowed to show the possibility of self-organization in open systems within the understanding of entropy as a measure of disorder. The application of the laws formulated for the entrostat has allowed to distinguish the fundamental cause of the global trends.

More than twenty years ago, precisely in 1990, the first article was published introducing the concept of the entrostat with regard to open systems [1]. At that time almost no one paid attention to the importance of this concept to explain the phenomenon of self-organization. In the recent years, at least in Russia, there has been a noticeable increase in the number of publications mentioning the term "entrostat". However, sometimes the authors do not use the term correctly. The impression is that they do not fully understand its meaning. In this connection, in this paper I would like to, first of all, recall the reasons why the concept of the entrostat was introduced in statistical physics; secondly, examine its features more fully; and thirdly, show the advantage that the researcher gets when using this term in the study of the systems' self-organization.

### 1. Concept of the Entrostat.

The concept of the entrostat appeared in order to resolve the crisis in statistical physics that called into question the interpretation of entropy as a measure of disorder. The main argument of the opponents of this interpretation was the reference to the fact that statistical physics failed to provide an explanation of the phenomenon of self-organization. The problem was only exacerbated by the apparent successes made in this direction by the part of synergetics based on the methods of nonlinear dynamics.

The main obstacle to the explanation of self-organization in statistical theory is the difficulty comparing the entropies of different states of an open system (see [2, 3]). What do I mean by this? If the system under study interacts with another system, changes occur in both of them. To describe these changes the appropriate variables are required. If these two systems constitute a general closed system, then in principle we can determine all the changes in each of them. Herewith, in the statistical expression of entropy (Boltzmann-Gibbs entropy) written for the system under study, the distribution function (the probability density) becomes conventional, depending on the variables of *both* systems. However, in most real cases the system under study is affected by a very large number of external systems - the external environment. In such cases, it is almost impossible to consider all the required variables and, therefore, it is impossible to record properly the statistical expression of entropy.

The introduction of the concept of the entrostat allowed to address the issue of comparing the entropies of different states from an unexpected viewpoint. As a result, the study of self-organization within the statistical theory made headway.

Entrostat is a system whose entropy doesn't change during the interaction with the system under study [1,4,5]. In practice, the property of an entrostat belongs to the environment for which the condition is fulfilled [6–9]:

$$\frac{|\Delta S|}{S} \gg \frac{|\Delta S_e|}{S_e},$$

where $\Delta S$ and $\Delta S_e$ are the entropy changes of the system under study and the external environment respectively, caused by their interaction.

As we can see, the external environment acts as the entrostat if its entropy change is neglectably small in comparison with the entropy change of the system under study. For example, in the tasks on thermal conductivity, the entrostat is represented by the thermostat – the environment that maintains constant temperature at the system's boundaries.

In the world around us, we can consider as the entrostat the environment whose influence upon the system greatly exceeds the reverse influence of the system upon the environment. For example, a strong wind blowing against the walking man forces him to lean forward, i.e., to perform certain actions. Since these actions do not affect the atmosphere, the atmosphere is the entrostat in relation to the man. Big noise outside an open window created by the traffic will force us to close the window, while the traffic – the entrostat – won't even notice this action. It is well known that it is easier to follow bureaucratic regulations than to procure their cancellation, i.e., the laws of society are the rules that one person interacting with the society (the entrostat) is almost unable to change. And such examples abound.

According to the author of the present paper, the phenomenon of the entrostat occurs when the impact of the system under study upon the entrostat is so small that it is comparable to disorderly structural noise in the entrostat (i.e., impact is infinitely small in relation to the entrostat).

The main advantage of introducing the concept of the entrostat is that it excludes the external environment when studying the behavior of an open system. In particular, all the changes that occur during the interaction of the system with the entrostat relate to the system. Therefore, all the new variables that describe these changes will relate to the system. The latter means that we do not need to build a general closed system. According to this, on the basis of the conventional entropy properties' analysis, the following relation was proved (the proof can be found in [5–8]):

$$S[X] > S[X|Y_1] > S[X|Y_1Y_2] > ... > S[X|Y_1Y_2...Y_i] > ... > 0, \qquad (1)$$

where the square brackets are the designation but are not the functional dependency; $X$ is the variable characterizing the state of the system;

$$S[X] = -k \int_X f(X) \ln f(X) dX$$

– the closed system's entropy in equilibrium state;

$$S[X|Y_1Y_2...Y_i] = -k \int_X \int_{Y_1} ... \int_{Y_i} f(XY_1Y_2...Y_i)\ln f(X|Y_1Y_2...Y_i)dXdY_1dY_2...dY_i \qquad (2)$$

– the entropy of the open system in *i*-th stationary state ( *i*-th stationary state differs from the closed state by the changes in the structure that arise due to the influence of the *entrostat* and the described variables $Y_1$, $Y_2$,..., $Y_i$ ); $f(X)$, $f(XY_1...Y_i)$, $f(X|Y_1...Y_i)$ are distribution functions (probability densities).

Note that expression (2) characterizes the system's entropy only if it interacts with the entrostat. Otherwise, for the distribution functions appearing in (2), it would be necessary, in addition to the variables $X, Y_1, Y_2,...,Y_i$, to take into account the variables that describe the changes in all systems with which the system under study presumably interacts. The latter, as noted above, is almost impossible.

In other words, in (1) the comparison of open states of the system is correct when we speak of the system's interaction with the entrostat. Neglecting this circumstance may lead to inconsistencies [7].

It is necessary to distinguish the situations in which neither of the two interacting systems can be considered the entrostat. Imagine that a hot object has been carried into a thermally insulated room. After some time the temperature of the object and the air temperature in the room become equal. Herewith, the change in temperature will be noticeable in both systems. Consequently, none of them can act as the entrostat. Now suppose that there is a window wide open in the room. After a while the hot object inevitably cools, its temperature will be exactly equal to the air temperature outside. As the temperature of the air remains the same after the object cools, in this case the air must be considered the entrostat.

Let us discuss some important consequences of inequality (1).

The entropies presented in (1) correspond to stationary states and differ in the number of variables. But, as noted above, the number of variables is determined by the value of external influence, i.e., the value of the system's *openness*. Therefore, for each inequality in (1) we shall assign a specific value of some phenomenological parameter called the system's degree of openness.

The *degree of openness* $\alpha$ is a parameter characterizing the value of all the structural changes that have occurred in the system as a result of its interaction with the entrostat [4,5,7].

The extreme positions of (1) hold the boundary states of the system. For the leftmost position $\alpha = 0$ is carried out, which means an absolutely closed state; for the extreme right position: $\alpha = \alpha_{max}$, which should mean a maximally open state.

Let us denote the entropy value in the *i*-th stationary state with a degree of openness $\alpha_i$ as $S_{\alpha_i}$, then (1) can be rewritten as follows:

$$S_{\alpha=0} > S_{\alpha_1} > S_{\alpha_2} > ... > S_{\alpha_i} > ... > 0$$

Graphically, this inequality can be expressed as the *entropy range* shown in Figure 1. In this figure: $S_0$ – the value of the system's entropy at the beginning of certain processes; $S_{AC} = S_{\alpha=0}$ – the entropy value at the end of these processes in an absolutely closed state of the system ($\alpha = 0$).

The entropy range is the graphic illustration of the regularity contained in inequality (1): if we increase the openness of the system from $\alpha_1$ to $\alpha_2$ its entropy must decrease from $S_{\alpha_1}$ to $S_{\alpha_2}$, i.e. the order will increase in the system, though not to infinity, but to a level corresponding to the new degree of openness; conversely, if we reduce the openness of the system from $\alpha_2$ to $\alpha_1$ its entropy must increase from $S_{\alpha_2}$ to $S_{\alpha_1}$, i.e. the disorder will increase to a level corresponding to the new degree of openness.

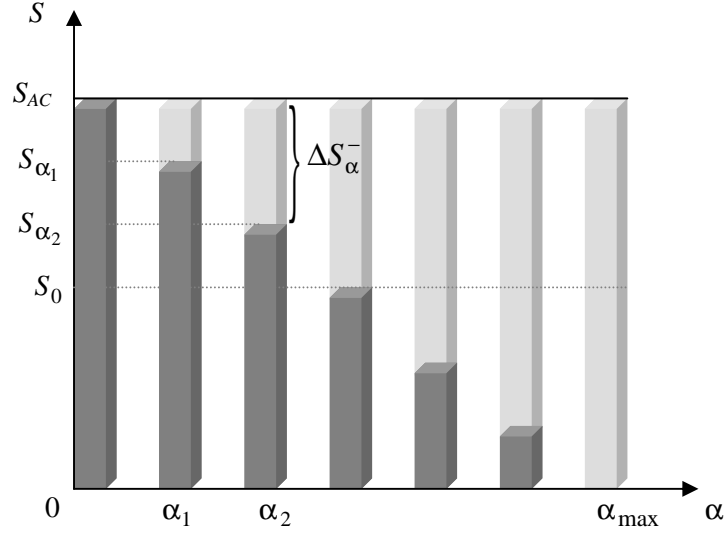

Figure1. Entropy range of the system – values of entropy $S$ for the stationary states of the system at its different degrees of openness $\alpha$ ranging from zero to maximal [5]. The shaded part of the column conditionally shows the entropy value that the system has in the stationary state at a certain value of the degree of openness.

Let us introduce the designations (see Figure 1):

$$\Delta S_\alpha = S_\alpha - S_0$$

– the change of the entropy of the system that has reached the stationary state (herewith system has a degree of openness $\alpha$);

$$\Delta S_{AC} = S_{AC} - S_0 > 0$$

– the entropy change of an absolutely closed system that has reached the equilibrium;

$$\Delta S_\alpha^- = S_\alpha - S_{AC} < 0$$

– the negative entropy change by which the stationary value $S_\alpha$ decreases with the increase of $\alpha$. It is easy to see that

$$\Delta S_\alpha = \Delta S_{AC} + \Delta S_\alpha^-.$$

We have found that in an open system the overall change of the entropy $\Delta S_\alpha$ consists of the positive $\Delta S_{AC}$ and the negative $\Delta S_\alpha^-$.

The value $\Delta S_\alpha^-$ was called *the critical level of the system's organization* [4–9]. In other words, we consider $\Delta S_\alpha^-$ as the quantitative measure of the organization of the system in the stationary state. According to (1), by the organization of the system we understand the structural connections described by the macroscopic variables $X$, $Y_1$, $Y_2$, ..., $Y_i$, ....

As seen from Figure 1, the value $|\Delta S_\alpha^-|$ uniquely corresponds to the system's degree of openness $\alpha$: $\alpha_i \Leftrightarrow |\Delta S_{\alpha_i}^-|$. This implies an important conclusion: to increase or decrease the value of the system's critical level of organization $|\Delta S_\alpha^-|$ it is necessary to correspondingly increase or decrease its degree of openness.

Summing up, let us formulate the laws of behavior of open systems under the influence of the entrostat (the detailed proof can be found in [5–8]):

1. Each open system has a critical level of organization. If the system is ordered below this level processes of self-organization prevail in it, if it is ordered above this level the processes of disorganization prevail; at the critical level these processes balance each other, and the state of the system becomes stationary.

2. The value of the critical level uniquely corresponds to the value of the influence of external environment (the entrostat) upon the system, i.e. the system's degree of openness.

3. To increase the order in the system, it is necessary to increase its degree of openness. The new value of the degree of openness will be matched by a new higher critical level of ordering. As a result, the system will be dominated by the processes of self-organization increasing the order to a new critical level.

4. To disorganize the system, it is necessary to reduce its degree of openness. Herewith, the critical level will fall, causing a predominance of disorganization processes that reduce the order in the system to a new value of the critical level.

There is a great number of facts in the surrounding world to illustrate these laws.

For example, in a closed vessel completely filled with some fluid, the motion of the molecules is equiprobable in all directions. Let us make a hole in the vessel. If the pressure in the vessel is higher than in the external environment, the molecules will quickly form a new structure $\tilde{\ }$ the fluid flow in the direction of the hole, i.e., self-organization takes place. To keep ourselves in a good physical shape, we execute various power exercises. In other words, we expose ourselves to external influence, causing self-organizing processes within the organism. The easing of physical activity inevitably leads to some disorganization of the organism. The living organism infected with the virus gets ill. But even in this case there are powerful processes of self-organization going on in its organism, as a result of which, firstly, it recovers, and secondly, becomes immune to the virus for a certain period. Of course, later, because of the absence of the virus in the environment, the organism will disorganize a bit, and will again be able to contract the disease. Medical vaccinations are in fact that small external influence that causes the organism to self-organize to a level sufficient to resist the virus.

What do parents do to reduce the disorder that constantly appears around the growing children? They educate the children, that is, they influence them creating in their consciousness the processes of self-organization. But once they get a little distracted from child education and let their kids become more reserved in relation to the parents, the processes of disorganization inevitably start dominating in them, and the result is what we call "getting out of hand".

Human society is also a system, so the described regularities exist in it. For example, without delving into the history of different states, you can still note that the states that tend to toughen the access control at their borders (a decrease in the degree of openness) experience within themselves an increase in the destructive processes in economy and culture, and in other areas of human activity that fall under the customs pressure. Conversely, the weakening of access control at the borders (an increase in the degree of openness) leads to strengthening progressive processes.

In [7, 10] there are other examples of the above-stated provisions.

Important Note: while opening the system with the aim of its ordering and self-organization one should ensure that the intensity of the opening does not exceed a certain threshold above which the system is not in time to self-organize, will be destroyed. For example, a military intervention is also an opening of the state that has been attacked. According to the above-stated regularities, powerful self-organization processes occur in this state: mobilization, intensification of all the productive forces, etc. However, if the intervention is fast and the state is small, it will not have the time to build effective defense and will be destroyed.

According to the well-known statistical expression of entropy, the latter is rigidly connected with the probability of events through the distribution function. Therefore, in practice the entropy laws manifest themselves in an *increase of the probability of the corresponding events*. In other words, the events contributing to these laws will occur *more frequently* than the others. But to what extent more frequently? At least, so that the requirements of the entropy law are fulfilled.

For example, in any system (for example, in the city), if we construct so much that it exceeds its critical level of system organization, the probability of events leading to disorganization will increase. In particular, speaking about the city, environmental problems will worsen, the number of accidents will increase, the number of causes for social unrest will grow, etc. Unfortunately, no major construction includes preliminary assessment of the entropy changes that may occur in the environment as a result of this construction.

Thus, the introduction of the concept of the entrostat has allowed to show the possibility of self-organization in open systems within the understanding of entropy as a measure of disorder.

**2. Global trends.**

The introduction of the concept of the entrostat allowed as well to find the answers to a number of issues that have recently become increasingly relevant for the human civilization. By transforming the world the man increases or decreases the order in it, that is, changes the entropy of the environment. What is the net change of entropy produced by the whole mankind: is it more or less than zero? The laws formulated in the previous section provide a definitive answer.

According to the Author, the cosmos in relation to the Earth can be considered as the entrostat. Consequently, the Earth has a degree of openness which for a very long period (at least longer than the lifetime of the human civilization) has been relatively constant. This degree of openness sets a certain critical level of our planet's organization $|\Delta S_\alpha^-|$. According to the above-stated laws, the processes of ordering and self-organization ($\Delta S < 0$) will prevail on the Earth below the critical level. The processes of disorganization ($\Delta S > 0$) will prevail above the critical level. In the first case the mankind, by transforming the world around us, increases more the order rather than the disorder. How long can this last? Until the moment when, in the process of creation, it exceeds the critical level of the planet's organization $\Delta S_\alpha^-$. In this case the processes of disorganization will prevail. As a result, the probability of destructive events will increase and the surplus built by the mankind beyond the critical level will be destroyed (or compensated by destructive events in the environment). By inertia, more than is sufficient will be destroyed to drop to the critical level. Below the critical level, the processes of self-organization will prevail and the humanity will again build houses and factories, other complex structures, that is, reduce the entropy of the environment. After some time it will again exceed the Earth's critical level. Then everything repeats: entropy oscillations arise (mathematical proof of the possibility of entropy oscillations can be found in [7, 9, 10]).

During the period of exceeding the critical level the entropy laws increase the probability of any events contributing to an increase of the disorder on the planet, i.e., form a destructive trend. What can be the sign of this trend? First, the increased intensity of natural disasters (hurricanes, earthquakes, floods, large fires etc.), the collapse of ecosystems (environmental crisis), the devastating climate variations. Second, the increased probability of technological disasters, accidents, wars. And third, any other destructive phenomena.

Note: all the three classes of phenomena have a common denominator – the increasing disorder. According to the author of the article, the latter means that these phenomena fall into the scope of the entropy laws described in this paper.

Understanding the reasons of the given trend not only allows to anticipate the coming ordeal but also suggests a way to circumvent it. As it was noted in the previous section, there is a univocal correspondence between the system's degree of openness $\alpha$ and the critical level of organization $|\Delta S_\alpha^-|$. In other words, by increasing $\alpha$, we shall increase $|\Delta S_\alpha^-|$. This will result in the processes of ordering and self-organization prevailing in the system. Thus, in order to make the processes of self-organization prevail in the system "Earth" it is necessary to increase the value of its critical level of organization. For this purpose it is necessary to increase its degree of openness (for example, as a result of a large-scale space exploration of cosmos – the Moon, the Mars, etc.). In result constructive trends will be formed on the Earth (i.e. the probability of events contributing to construction will increase), and the intensity of destructive trends will diminish. It should be understood that the entropy laws described here leave us with only one option: the world we live in has to be opened constantly. Herewith, each time the mankind would delay the next large-scale opening, the threat of total destruction would arise again [4, 9, 11]).

## 3. Summary.

1. The external system whose entropy change can be neglected in comparison with the entropy change of the system under study acts as the entrostat. The fundamental reason for this is the fact that the impact of the system under study upon the entrostat doesn't exceed the disorderly structural noise in the entrostat, i.e., this impact represents an infinitely small value for the entrostat.

According to the author of the present paper, when the researcher refers to the entrostat he or she has to deal with transition from one level of description to another, i.e., with the transition through singularity. For example, in hydrodynamics, an approximation of continuum implies that the element of the continuum is the point, which in its turn consists (and this is considered crucial) of a large number of molecules of liquid. In this case no one would think it necessary to consider each molecule when writing the equation of the fluid motion. This is not the right level of description. In other words, taking away or adding one of the molecules makes absolutely no effect on the behavior of the continuum. The reason for this may be only one – the continuum is an infinitely large system for a single molecule, i.e., it is the entrostat.

2. The introduction of the concept of the entrostat has allowed to show the possibility of self-organization in open systems within the statistical understanding of entropy as a measure of disorder.

3. The application of the laws formulated for the entrostat to the system "Earth" has allowed to distinguish the fundamental cause of the global trends. A relatively constant degree of openness of the Earth (in relation to the cosmos) sets a certain critical level of the order on the planet. Below its critical level, the mankind increases the order in the environment more that the disorder, thus it inevitably aims at exceeding this level. When it happens processes of disorganization should prevail on the Earth, i.e., the probability of any events leading to distruction should increase. At the same time, increasing the planet's degree of openness (for example, as a result of large-scale cosmos exploration – of the Moon, the Mars, etc.) would increase the value of its critical level. The latter will form strong creative trends.